\documentclass[MSNbibl,number,dvips]{arxstspdf}
\usepackage{flushend}
\usepackage{stfloats}

\volume{27}
\issue{3}
\pubyear{2012}
\firstpage{344}
\lastpage{345}
\doi{10.1214/12-STS381B} 
\referstodoi{10.1214/11-STS381}

\begin{document}
\begin{frontmatter}
\vspace*{6pt}
\title{Discussion of ``Multivariate Bayesian Logistic
Regression for Analysis of Clinical Trial Safety Issues''
by W.~DuMouchel}
\runtitle{Discussion.}

\begin{aug}
\author[a]{\fnms{Don} \snm{Berry}\corref{}\ead[label=e1]{dberry@mdanderson.org}}
\runauthor{D. Berry}

\affiliation{University of Texas M.D. Anderson Cancer Center}

\address[a]{Donald A. Berry is Professor, Department of Biostatistics, University of Texas M.D. Anderson Cancer Center, Houston, Texas
77030-1402, USA}

\end{aug}



\end{frontmatter}

Drug safety is a major medical concern. Appropriately so. And recent
high profile cases have heightened the level of concern. These cases
include\break Vioxx~\cite{r1}, Vytorin and its components \cite{r2}, Tysabri~\cite{r3} and
Avandia \cite{r4}.

These and other cases and the controversies they have engendered have
increased awareness that dealing with and understanding drug safety
issues is enormously difficult. No doubt the final verdict has been
wrong in some cases (not necessarily any of the ones mentioned above).
Drugs are not protected by the ``innocent until proven guilty''
principle. Just as with national security measures, heightened awareness
is good, but overreaction can be detrimental to delivering good
medicine.

Inferential problems related to drug safety are numerous as well as
difficult. First, there are many types of serious adverse effects to
consider. Drugs can kill or induce potentially fatal conditions. They
can also lead to one or more effects that detract from the patient's
quality of life. Multiplicities abound. Moreover, the same effects
usually occur naturally, perhaps even as part of the disease process for
which the drug is being used. The statistical question is whether and
which serious adverse effects occur at an increased rate for patients
taking the drug.

The medical questions are also difficult. All drugs cause some side
effects, usually in a dose-dependent fashion. So the issue is the
benefit/risk trade-off. For example, the same serious adverse effect can
have a very different implication in treating cancer, say, than in
the\vadjust{\goodbreak}
primary prevention of cardiovascular events. Indeed, for some cancer
therapies, certain adverse effects are a good thing because they
indicate that the therapy is doing a better job of fighting the tumor:
``Congratulations, Ms. Smith, your hot flashes mean the drug is
working!''

Compounding the multiplicity of types of adverse effects is the
multiplicity of drugs, their doses, and combinations. For any particular
adverse effect upon which no drug has an impact, the data will show that
half of the drugs have some amount of increase in the incidence of that
effect. And some of these increases will be statistically significant. A
small proportion of the drugs will be shown to be detrimental
statistically in any particular comparison, but there are many
comparisons. How to separate the signal from the noise? And how to
balance false positives (rejected drugs that are safe) with false
negatives?

Bonferroni and other traditional adjustments for multiple comparisons
are inappropriate when the measurements concern safety (and they may
never be appropriate!). They are used to protect against rejecting too
many null hypotheses. When the question is one of safety, this would
mean the more comparisons one makes, the more difficult it is to
determine that a drug is unsafe.

Bill DuMouchel has a long history of developing and using Bayesian
hierarchical modeling methods for addressing multiplicity problems
associated with large, sparse databases. His data mining approaches as
applied to questions of drug safety have been used by the U.S. Food and
Drug Administration, among others. The methodology he has previously
developed gives a clear view through muddy waters. His article in this
issue makes the view even clearer.

DuMouchel's application of multivariate Bayesian logistic regression
(MBLR) ``borrows strength'' in the usual Bayesian hierarchical modeling
sense. For example, if a drug seems to increase both nausea and
vomiting, then the conclusion about both adverse effects is stronger
than if either were considered by itself. On the other hand, if the
incidence of nausea is elevated but that of vomiting is not, then any
conclusion about nausea based on all the evidence is less compelling.
The borrowing is across clinical trials as well as across related side
effects.

Neither aspect of the borrowing in MBLR is novel on its own. Bayesian
hierarchical modeling is a standard approach to meta-analysis. And
borrowing hierarchically across side effects within body systems has
been proposed previously \cite{r5}. But putting the two together is novel. And
it is an important concept. There are usually many clinical trials
conducted of a drug, most with the primary focus on efficacy. It is
important to take advantage of all the evidence. Safety applies to the
drug and not to the trial. A~safety signal may be observable only over
several trials. DuMouchel's methodology is consistent with the synthetic
nature of the Bayesian approach.\looseness=1

This elegant article with its methodology is a welcome addition to this
important problem area. MBLR will become a standard method for
determining\break whether a drug increases the incidence of any adverse drug
effects. It will also provide appropriate estimates for any
increases.\vspace*{3pt}


\end{document}